# Biological detail and graph structure in network neuroscience

David Papo[1,2,*], Javier M. Buldú[3]

[1] *Department of Neuroscience and Rehabilitation, Section of Physiology, University of Ferrara, Ferrara (Italy)*
[2] *Center for Translational Neurophysiology, Fondazione Istituto Italiano di Tecnologia, Ferrara (Italy)*
[3] *Complex Systems Group & GISC, Universidad Rey Juan Carlos, Madrid (Spain)*
[*] *Email address: david.papo@unife.it*

**Abstract**

Representing the brain as a complex network typically involves approximations of both biological detail and network structure. Here, we discuss the sort of biological detail that may improve network models of brain activity and, conversely, how standard network structure may be refined to more directly address additional neural properties. It is argued that generalised structures face the same fundamental issues related to intrinsicality, universality and functional meaningfulness of standard network models. Ultimately finding the appropriate level of biological and network detail will require understanding how given network structure can perform specific functions, but also a better characterisation of neurophysiological stylised facts and of the structure-dynamics-function relationship.

## 1. Introduction

It is intuitive to represent brain anatomy and the activity that it produces as a network structure, i.e. as a collection of nodes and connecting edges (Bullmore and Sporns, 2009), and ultimately to study the role of this structure in brain dynamics and function (Papo et al., 2014b; Papo and Buldú, 2025b). A network structure provides a compact, inherently multiscale, characterisation of multi-body systems, possibly preserving its intrinsic properties and symmetries. Moreover, network structure can affect on network dynamics and the processes unfolding on it (Boccaletti et al., 2006; Masuda et al., 2017) and can interact in complex ways with local dynamics (Gross and Blasius, 2008). Thus, network structure may to some extent explain brain dynamics and function, and may help predicting the system's behaviour, quantifying its evolvability, and, at least in principle, controlling it (Liu and Barabási, 2016), or steering it to desired states (Gutiérrez et al., 2012, 2020).

Complex network theory is a statistical mechanics approach to graph theory (Albert and Barabási, 2002). In this approach, justified by the sheer number of components (Chow and Karimipanah, 2019), the identity of nodes and links loses importance, at least *prima facie*, as the network's properties are statistical in nature. Implicit in a statistical mechanics' approach is the fact that seemingly profoundly different physical systems may be characterised by the same collective behaviour which can be grouped in universality classes. The large scale behaviour of each class can be described in terms of simple effective models specified in terms of an interaction network and a limited number of control parameters, where only a small number of relevant features, *viz.* symmetries, dimensions, and conservation laws, turn out to be relevant, while microscopic details can be disregarded. But, how universal are brain network representations? To what extent and at what scales do brain dynamics and function depend on the specific details of their nodes and links?

We review aspects of neural activity that may be incorporated into neural network modelling and somehow dually, network models that may help representing brain structure, dynamics, and function. We then address the following two-fold question: what neural and network properties should be incorporated for a network structure to reproduce known anatomical patterns and dynamical phenomenology and to allow a faithful representation of functional brain activity?

## 2. Ground level of brain network modelling

In its most general form, a network is a structure $\mathcal{N} = (V, E)$, where $V$ is a finite set of *nodes* or *vertices* and $E \subseteq V \otimes V$ a set of pairs of *links* or *edges* $V$. The links can carry a *weight*, parametrising interactions' strength, and a *direction*. All the information in a network structure is contained in the associated connectivity matrix. encoded into its combinatorial (Bollobás, 1986), topological, and geometric properties (Boccaletti et al., 2006), and its symmetries (Dobson et al., 2022) (*See* A1).

In real space, the microscopic scale may be identified with neurons, or neuronal masses at various scales, and may contain more or less biological detail. Cortical columns are often treated as cortical systems's basic dynamical units, which are coupled through sparse long-range cortical connectivity. Thus, at system-level, neocortical activity is often modeled as an array of weakly-coupled dynamical units, whose behaviour is represented by dynamical attractors of various types (Breakspear and Terry, 2002) (*See* A2). In its simplest form, the system's units are static. The system's units can also be thought of as dynamical systems (Golubitsky and Stewart, 2002a), e.g. spiking neurons and the resulting system is a discrete-space, continuous-time dynamical system (DeVille and Lerman, 2015). Thus, overall, a neural system can be thought of as a set of dynamical systems, whose state variables evolve e.g. according to differential equations and whose interactions are encoded by a graph (Bick et al., 2023). The state of a system can also be defined by the time-varying interplay between its structure and the variable's dynamics unfolding on it (Ghavasieh and De Domenico, 2022).

Irrespective of the context and the space in which a network structure is defined, the neurophysiology-network representation map often involves drastic simplifications on both sides of the map. For instance, a great number of studies, particularly at macroscopic scales, are predicated upon a *simple network* structure. A network is said to be *simple* if it has neither self nor multiple edges between the same pair of nodes (in the same direction for directed networks). In spite of its apparent generality, some known anatomical and dynamical neural stylised facts are not accommodated within the simplified structure used in these studies and this may in principle limit the ability to account for known phenomenology or to reveal as yet unknown one.

In the remainder, we consider a ground level of network structure and use its underlying assumptions and corresponding limitations to analyse on the one hand the neural aspects of brain activity that are not easily accommodated by such a structure and, on the other hand, the nature of the possible network structure that could better reflect intrinsic properties of brain structure, dynamics, and function.

## 3. Adding biological detail

Both theoretical and experimentally derived network representations typically drastically simplify details of actual brain anatomical and dynamical structure at all scales, including that of single cells. For instance, standard network representations do not include features of neural activity such as hardware heterogeneity, recurrence, or inhibition or, when modelling long-distance inter-areal pathways, the laminar and anisotropic character of the connections (Markov et al., 2013) in addition to their strength and specificity or and the resistive nature of brain tissue. A number of questions ensue: how much and what sort of detail should be added and at what scales? How would refining neurophysiological information change brain models?

### 3.1. Nodal properties

Various aspects of neural activity are in general thought of as reducible to network nodes. The anatomo-functional criteria allowing this reduction are scale-dependent, the most obvious aspects being related to the cell-level structure of the brain. At neuronal scales, such reduction typically involves various simplifying assumptions on synaptic structure and physiology, including assumptions on hardware, viz. on its homogeneity or, more generally, on the physical units responsible for brain dynamics and function homogeneity, but also on the way afferent information is integrated to produce cell firing.

*3.1.1. Defining meaningful neural units*

A network representation requires identifying meaningful neurophysiological units (Korhonen et al., 2021). Though *prima facie* straightforward, this step is nonetheless non-trivial, even at the single neuron scale. Indeed, activity at subneural scale can be related to function at macroscopic scales (Li et al., 2024). Moreover, achieving an appropriate characterisation that captures the essence of neuron information processing activities requires defining independent electrical processing units explaining its overall input–output behaviour (Koch et al., 1982). Although dendrite arborisations and axon terminals already present a network structure carrying out computationally complex operations (Gidon and Segev, 2012), single neurons are often thought of as simple point-like units, where all synapses have an equal opportunity to influence neuronal output, and the output results from a linear weighted sum of all excitatory and inhibitory synaptic inputs. However, pyramidal cells' terminal branches of the apical and basal trees constitute sets of independent non-linear subunits (Häusser and Mel, 2003). In general, one can distinguish separate functional compartments in the dendritic tree, the number of which depends on the considered aspect of dendritic function, based on the effects that such compartments and their interactions exert on the neuron's computational power and synaptic plasticity. The spatial extent of propagation of the dendritic spike will also define the spatial range of plasticity. Functional compartments can be defined at scales even finer than those of thin branches. Specifically, the rules for induction of synaptic plasticity may differ at proximal and distal synapses in a way that is defined by the properties of their respective compartments.

The issue is replicated at coarser scales in real space as well as in phase space, as finding meaningful criteria for separation and discretisation becomes more challenging.

*3.1.2. Hardware heterogeneity*

Both excitatory and inhibitory neurons come in a large number of different types which differentially affect cross-variability, both by their specific connectivity and by their intrinsic properties (Balasubramanian, 2015). However, at a given scale, particularly when considering static network structure, all network nodes are typically assumed to be essentially identical. This approximation may be acceptable at certain scales, but perhaps not at others, particularly at the whole system level, and may serve certain goals, e.g. estimating information transport via degree distribution, but may be misleading whenever function is not an emerging property of topology, e.g. at scales at which information processing is done at nodal scales (Sterling and Laughlin, 2015).

An important question is how node heterogeneity, e.g. in excitability or in coupling strength, may affect collective dynamics. Heterogeneity in excitability across units may play a double role: during states of low modulatory drive, it enriches the system's dynamical repertoire; on the other hand, it acts as an effective homeostatic control mechanism by damping responses to modulatory inputs and limiting firing rate correlations, ultimately decreasing in a context-dependent way the system's susceptibility to critical dynamical transitions (Hutt et al., 2023; Balbinot et al., 2025).

Neural heterogeneity may also play a role in neural networks' computations (Gast et al., 2024). If neural systems' information-processing capabilities are related to the morphological diversity of neurons, a reliable description of neuronal morphology should be key to the characterisation of neural function, although what level of detail is would be necessary and sufficient to determine function remains to be determined. Note, though, that while morphological information may be thought of as a proxy for function, it does not constitute a necessary or sufficient condition for it.

Finally, an important issue is whether a statistical mechanics is possible given the number of qualitatively different pieces of hardware. Microfoundations of models would imply a detailed description of the hardware. This may seem to weaken the pillars of the statistical mechanics approach underlying graph theoretical modelling, *viz*. a loss of important symmetries (exchangeability, scaling, and universality).

*3.1.3. Beyond neurons*

An important question is whether brain dynamics can be understood just in terms of classic excitable units, i.e. neurons, or other units. For instance, in the human brain, glia cells are approximately as numerous as neurons and are tightly integrated into neural networks (Herculano-Houzel, 2014) but are in general not accounted for in brain network models (Turkheimer et al., 2025). Glial cells play a key role in the development of vascular and neural networks and control homeostatic processes in the mature brain, provide neurons with energy, supply neurons with neurotransmitter precursors and catabolise neurotransmitters (Verkhratsky and Nedergaard, 2018). In particular, astrocytes are key to fundamental processes in brain networks' building, dynamics and repair, regulate synaptic maturation, maintenance, and extinction, and play an important role in the orchestration of synaptic plasticity (De Pittà and Berry, 2019) and in the restoration of connectivity and synchronisation in dysfunctional circuits, e.g. in cerebellar networks (Kanner et al., 2018). Astrocytes actively communicate with neurons, through a process termed *gliotransmission* (Araque et al., 2014). While their exact mechanisms and functions are poorly understood, gliotransmitters activate neuronal receptors and account for astrocyte-mediated modulation of synaptic transmission and plasticity (Savtchouk and Volterra, 2018), acting as spatio-temporal integrators, decoding information in large arrays of neuronal activity. The relationship between neocortical neurons and astrocytes is a critical factor determining the effects of endogenous and exogenous electric and magnetic field interactions (Martinez-Banaclocha, 2018). For example, while seizure discharges ultimately result from neuronal activity, glias may play an important role in excitation and inflammation in seizures kindling and modulation (Devinsky et al., 2013). More generally, atypical neuron-glia interactions are implicated in brain pathology, viz. in schizophrenia (Radulescu et al., 2025). Finally, synapse-astrocyte communication may also play a fundamental role in cognitive function, e.g. in working memory (De Pittà and Brunel, 2022).

### 3.2. Link-related properties

Loss of neurophysiological detail in network modelling is also found at the level of bare connectivity. This is partly due to simplification of the anatomical connectivity structure to accommodate it to that of a simple network and partly to lack of knowledge of the functional mechanisms of neural information transport and computation.

*3.2.1. Wire properties*

When considering neural systems in real space, links represent brain fibres at all scales, and of interest is how these structures support activity. The amount of current or information conveyed by a link depends on wires' physical characteristics, such as their diameter and length but also their mechanical and conduction properties (Sterling and Laughlin, 2015). Wire geometry therefore contains important information at time scales ranging from evolutionary to experimental.

An important neural property often not incorporated in graph theoretical models of brain activity is *load*, a local measure given by the ratio between flow and capacity. Together with network topology, information on load and its distribution may be crucial in the prediction of link failure on network processes and to understand which links are critical to a given function (Witthaut et al., 2016).

*3.2.2. Delays*

Brain networks are embedded in the anatomical space and this leads to time-delays due to finite signal propagation speed. Time-continuous delay systems, which exhibit in practice high dimensionality and short-term memoryexpress a variety of dynamical regimes, ranging from periodic and quasiperiodic oscillations to deterministic chaos (Ikeda and Matsumoto, 1987). Delays can facilitate zero-lag in-phase synchronisation (Ernst et al., 1995; Atay et al., 2004; Fischer et al., 2006) and can both stabilise and destabilise dynamical systems (Schöll and Schuster, 2008). Moreover, delay systems afford simple dynamical systems high-level information-processing capabilities (Appeltant et al., 2011).

Distance-dependent conduction delays are a crucial factor shaping brain dynamics and have a significant impact on the architecture of neocortical phase synchronisation networks (Deco et al., 2009; Petkoski et al., 2016; Roberts et al., 2019; Petkoski and Jirsa, 2019, 2022; Williams et al., 2023), inducing qualitative changes in the phase space of spatially-embedded networks (Voges and Perrinet, 2010). While topology can be thought of as a control parameter steering the dynamics through phase transitions, the dynamics is largely due to heterogeneous connectivity's time-delay, rather than changes in the topology (Jirsa and Kelso, 2000; Pinder et al., 2024). In the presence of delays, limit-cycle oscillators lead to collective metastable synchronous oscillatory modes at frequencies slower than the oscillators' natural frequency (Cabral et al. 2022). Time-delays also play an important role in neural networks' pattern formation (Muller et al., 2016; Roberts et al., 2019; Petkoski and Jirsa, 2022). For instance, spontaneous travelling waves may be an emergent property of horizontal fibre time delays travelling through locally asynchronous states (Davis et al., 2021). Moreover, in the presence of conduction delays, spike-timing dependent plasticity can exert activity-dependent effects on network synchrony in recurrent networks (Lubenov and Siapas, 2009). Finally, conduction delays are essential in long-range communication through coherence in the brain (Bastos et al., 2015).

*3.2.3. Activity propagation and flow directionality*

According to the *law of dynamic polarisation* (Ramón y Cajal, 1911), information unidirectionally flows from dendrites to soma to axon. However, for many types of neurons, excitable ionic dendritic currents allow dendritic action potentials traveling in the opposite direction (Stuart et al., 1997). Thus, the neuron itself contains an endogenous feedback mechanism. Backpropagating action potentials have many important consequences for dendritic function and synaptic plasticity (Linden, 1999). For example, a somatic action potential can trigger a burst due to its interaction with the dendrites (Häusser and Mel, 2003). Moreover, dendritic geometry, together with channel densities and properties, plays a crucial role in determining both forward and backpropagation of action potentials and dendritic spikes (Vetter et al., 2001). Likewise, synapses can propagate activity centrifugally but also centripetally, distributing input and output over the entire group of dendrites (Pribram, 1999).

*3.2.4. Connectivity density and anatomo-functional structure*

Both anatomical and dynamical brain networks have long been thought of as highly *sparse*. However, no general consensus exists over global estimates of brain activity. For instance, while estimates of the absolute number of axons suggested that human cortical areas are sparsely connected (Rosen and Halgren, 2022; Hilgetag and Zikopoulos, 2022) cortical areas may be far more connected than previously acknowledged (Markov et al., 2011; Wang and Kennedy, 2016). While in random networks, sparsity would ensure that neurons share a negligible proportion of presynaptic neighbours and inputs, and as a result that their activity is in general uncorrelated, this would not be the case in non-trivial, densely connected cortical populations (Pretel et al., 2024).

Densification may induce non-trivial structural transitions, including phase transitions in the scaling of the number of cliques of various orders with the number of network nodes and absence of self-averaging (Lambiotte et al., 2016), connectivity density may in principle affect network resilience, although neither anatomical disruption nor decreased connectivity are necessary conditions for functional disruption (Papo and Buldú, 2025a). From a modelling viewpoint, an incorrect density estimate, tantamount to downsampling the system (Wilting and Priesemann, 2018) may ultimately lead to underestimating network size. Near a phase transition, where correlations diverge, such systems this may lead to finite size effects, which can hide criticality or rare region effects. Moreover, a correct estimate of connectivity is key to obtaining a faithful representation of the associated dynamic patterns' dimensionality (Recanatesi et al., 2019). Moreover, while strong links may incorporate fundamental features of the system, weak links, often missed, particularly in experimental data analysis, may be needed to identify the system (Zanin et al., 2021, 2022), and failure to include them may lead to incorrect conclusions on network stability and robustness to network dismantling (Csermely, 2004).

*3.2.5. Mesoscopic structural principles*

Any model of brain cortical structure should incorporate or account for general organisational properties of its anatomy and physiology. For instance, the cerebral cortex exhibits a layered organisation, with the number of layers varying across phylogenetically different cortices. Moreover, various cortical and subcortical regions have a topographic arrangement, whereby spatially adjacent stimuli are represented in adjacent cortical locations, as well as a columnar structure whereby neurons within a vertical column share similar functions and connections and are connected horizontally to constitute functional maps (Hoffman, 1989; Mendoza-Halliday et al., 2024).

In almost all cortical areas, a substantial part of the output targets its area of origin (Douglas and Martin, 2007; Barak, 2017). In recurrent structures, a given neuron can receive input from any other neuron in the network, blurring the concept of upstream or downstream activity, so that their activity is affected the network's and not only by exogenous afferent input. Such a structural property enables networks to perfom computations at time scales much larger than those of a single stimulus, e.g. working memory, decision-making (Douglas and Martin, 2007), recall through pattern completion (Marr, 1971; Treves and Rolls, 1992), or integration of sensory information with stored knowledge (Singer, 2021).

### 3.3. The node-link contact area

Perhaps the most overlooked aspect on brain network modelling is the node-link junction. Type of contact and location are typically stylised, even at neuronal level. Furthermore, in large-scale models, the effects of contacts are modelled as flows, and therefore implicitly thought of as excitatory.

*3.3.1. Contact area and signal integration*

Both anatomically and functionally, the area through which different brain units contact other is sometimes difficult to characterise, even at the single neuron level. On the one hand, many neurons do not connect via linear one-to-one connections, but form neurites with collaterals, or branches at distinct segments of the main axon, connecting with multiple synaptic targets or highly branched synaptic termination zones (Spead and Poulain, 2020). On the other hand, action potentials are in general thought to be initiated in a particular subregion of the axon along which they propagate promoting neurotransmitter release at synaptic terminals. However, in some cases, neurons may be morphologically and dynamically different, e.g. they may not have a genuine axon, and the cell's basic functional aspects are undertaken by dendrites (Goaillard et al., 2020). Spikes can also be generated at dendrites, though their functional meaning is still poorly understood (Larkum et al., 2022). Furthermore, dendritic trees are often thought of as spatially extended systems consisting of passive cables, and electric current's spreading is understood in terms of cable equations, but signal integration rules within such a system, how they influence synaptic input processing, interact with different forms of plasticity, and ultimately contribute to the brain's computational power are still poorly known matters (Häusser and Mel, 2003). Moreover, evidence for the role of astrocytes in synaptic integration and processing, and for tripartite synapses, a configuration wherein astrocytes and neurons communicate bidirectionally (Perea et al., 2009), further complexifies contact area's functional structure at single neuron scales. Finally, contact areas are more complex to delineate at meso- and macroscopic scales, where both node contours and links' definition require context-dependent assumptions (Korhonen et al., 2021).

*3.3.2. Inhibition*

A key aspect of neural activity whose relationship with network structure remains difficult to incorporate is *inhibition*. Inhibition plays important roles at essentially all neural scales (*see* A3). At the single neuron scale, inhibitory inputs from distinct sources target specific dendritic subdomains, from distal to proximal dendritic regions (Markram et al., 2004; Jadi et al., 2012). This region-specific targeting plays a key role in controlling dendritic processes (Larkum et al., 1999; Isaacson and Scanziani, 2011), in synchronising their activity (Vierling-Claassen et al., 2010), and in regulating plasticity (Sjöström et al., 2008). Moreover, while excitation and inhibition are not symmetric in the way they compete for spike generation, inhibitory synapses are associated with high information transfer between spike trains, which are usually exclusively ascribed to excitatory synapses. At meso- and macroscopic scales, inhibition plays a crucial role in synchronisation of neural systems (van Vreeswijk et al., 1994). Inhibitory control of excitatory loops (Bonifazi et al., 2009) constitutes a generic organisational principle of cortical functioning, which stabilises brain activity (Griffith, 1963). For instance, inhibitory feedback can decorrelate a network (Tetzlaff et al., 2012). Moreover, inhibitory neurons have been proposed to play an important role in controlling the cortical microconnectome (Kajiwara et al., 2021). On the other hand, while evidence suggests that excitatory neurons form networks with non-trivial structure, whose fine-scale specificity is determined by inhibitory cell type and connectivity (Yoshimura and Callaway, 2005), inhibitory interneuron connectivity tends to be locally all-to-all (Fino and Yuste, 2011).

Neurons' collective dynamical regime, known as an *asynchronous state* (Renart et al., 2010), characterised by sporadic relatively uncorrelated firing with high temporal variability results from the interplay between excitatory and inhibitory forces (van Vreeswijk and Sompolinsky, 1996, 1998). Notably, such a balance relies on the role of glial cells, particularly astrocytes (Turkheimer et al., 2025).

Inhibition also constitutes an important ingredient for high-precision computation. The maintenance of an excitatory/inhibitory balance may allow cortical neurons to construct high-dimensional population codes and learn complex functions of their inputs through a spatially-extended mechanism far more precise than local Poisson rate codes (Denève and Machens, 2016).

How does inhibition affect network-related properties and the processes taking place on the network structure? First, inhibition plays an important role in routing (Wang and Yang, 2018). Second, it may affect network structure via plasticity mechanisms. In particular, interneurons contribute to the induction of long-term plasticity at excitatory synapses (Wigstrom and Gustafsson, 1985); conversely, excitatory transmission modulates inhibitory synaptic plasticity (Belan and Kostyuk, 2002). By modulating plasticity, inhibition, inhibitory plasticity and connectivity play important functional roles (Pulvermuller et al., 2021). For instance, inhibition controls the duration of sharp-wave ripples in hippocampal recurrent networks, which mediate learning (Vancura et al., 2023), while inhibitory plasticity supports replay generalisation in the hippocampus (Liao et al., 2024). Furthermore, inhibitory connectivity determines the shape of excitatory plasticity networks (Mongillo et al., 2018). On the other hand, while neural structure heterogeneity may locally affect the excitation/inhibition balance, the balanced state may be recovered through homeostatic mechanisms, which may themselves be regulated by inhibitory mechanisms (Pretel et al., 2024). Likewise, it has recently been shown that networks adapt to chronic alterations of excitatory-inhibitory compositions by balancing connectivity between these activities (Sukenik et al., 2021).

### 3.4. Multiscale and field-related properties

Up until now, we mentioned neural mechanisms which can be mapped onto particular regions of a network structure. However, other important neural mechanisms are not easily mapped onto local network structure. Arguably the two most prominent are neural mechanisms related to learning and adaptation and neuromodulation.

*3.4.1. Learning, plasticity, and adaptation*

Up until now, the focus has been on spatially local static or steady-state properties of neural activity. However, neural populations are able to change their properties in order to learn and adapt to new challenges from the environment. One important mechanism of brain plasticity is represented *Hebbian learning*, whereby the strength of connections between neurons increases when they are simultaneously activated (Hebb, 1949). Hebbian learning alone would lead to dynamic instability and runway excitation (Markram et al., 1997), and ultimately to complete circuit synchronisation (Zenke et al., 2017). Dynamic stability can be achieved in various ways, e.g. via *homeostatic plasticity*, through which neurons control their own excitability, ultimately regulating spike rates or stabilising network dynamics at various time scales (Turrigiano et al., 1998; Cirelli, 2017). Homeostasis can be implemented by various neurophysiological mechanisms, e.g. as synaptic scaling or efficacy redistribution (Turrigiano et al., 1998), membrane excitability adaptation (Davis, 2006; Pozo and Goda, 2010), or neuron-glial interactions (de Pittà et al., 2016). Synaptic plasticity may occur not only at synapses active during induction, but also at synapses not active during the induction. While these two mechanisms operate on the same time scales they have different computational properties and functional roles. The former mediates associative modifications of synaptic weights, while the latter counteracts runaway excitation associated with Hebbian plasticity and balances synaptic changes (Chistiakova et al., 2014).

Synaptic strength adjustment is only one among various possible homeostatic regulation mechanisms. A critical role in learning may also be played by suprathreshold activation of neurons and their integration. Neuronal activity is determined by excitatory and inhibitory synaptic input strength but also by intrinsic firing properties, which are regulated by the balance of inward and outward voltage-dependent conductances, respectively stabilising average neuronal firing rates and controling shifts between synaptic input and firing rate (Turrigiano et al., 1998).

Plasticity has been associated with the generation of complex dynamical regimes in recurrent neural networks. For example, synaptic facilitation and depression promote regular and irregular network dynamics (Tsodyks et al., 1998). Plasticity at inhibitory synapses can stabilise irregular dynamics (Vogels et al., 2011), while synaptic plasticity based either on activity strength (de Arcangelis et al., 2006; Levina et al., 2007, 2009) or on spike timing (Meisel and Gross, 2009; Rubinov et al., 2011) can induce critical fluctuations and phase transitions from random subcritical to ordered supercritical dynamics (Rubinov et al., 2011). Although often thought of as a purely local phenomenon, which would therefore be best understood as pertaining to node-link contact area, there are still considerable knowledge gaps regarding the spatial and temporal scale at which Hebbian, homeostatic and other plasticity mechanisms actually interact as well as their exact functional role (Wen and Turrigiano, 2024).

*3.4.2. Neuromodulation*

Neuronal activity is regulated at various spatial and temporal scales by numerous chemical messengers, including neurotransmitters, neuromodulators, hormones. These systems are often thought of a pointwise as they originate in well-defined brainstem and forebrain nuclei, and their effect is studied as a generic perturbation of neuronal network. However, one should distinguish between the quasi pointwise structure of neuromodulatory nuclei and the network-like structure of neuromodulation's consequences. these chemical messengers exert their effects through complex networks of diverging and converging pathways. For instance, different transmitters can act through the same network. Moreover, the effects of transmitters often depend on the presence of other transmitters and are characterised by higher-order functional phenomena such as *metamodulation*, whereby a modulator's action is gated by that of another modulator (Katz and Edwards, 1999). Of interest are then not only the effects of each of these chemicals on the topological properties of the neural network, but also those of the complex high-order network of neuromodulators. How does global network dynamics and functional space result from the multidimensional input space of transmitters? Should neuromodulation be thought of as an extrinsic structure? Does it have a network structure of its own or should it be considered as a modulator of a system it is not part of? If so, how should this interaction be modelled?

Neuromodulators have long been known to shape neural circuits (Bargmann, 2012). More specifically, it has been proposed that neuromodulatory systems enable the brain to flexibly shift network topology (Shine et al., 2019, 2021) in a state and activity-dependent way (Ito and Schuman, 2008; Sakurai and Katz, 2009). However, whether and how various neuromodulatory systems interact with plasticity mechanisms to facilitate brain function is poorly understood. In particular, on what type of network, what network property, how and at what scales do neuromodulators act?

## 4. Fine-tuning network structure

In the previous part, we examined some neural properties that are often not included in neural network modelling particularly at system-level scales. There is no clear picture of the information lost by network models simplifying brain structure and dynamics and, conversely, of the extent to which such network representations and the phenomenology that they may produce are robust to detail simplification.

inhere we examine network structure that could explicitly incorporate and account for key neural properties. Understanding the brain as a networked system has at least two important conceptual aspects. Equipping a system with a network structure comes with a number of assumptions and corresponding limitations. The conditions for reducibility to network structure, including, discretisability, intrinsicality, structure preservation have been discussed at length elsewhere (Korhonen et al., 2021; Papo and Buldú, 2024). We provisionally assume that the system can adequately be described as a networked system at least at some level, but that the network structure used to model such a system may fail to incorporate important aspects of its anatomy, dynamics and physiology. This implies that network structure is "relevant" to some important aspect of the system, in particular to its dynamics and function. Conversely, understanding the brain as a particular network structure has implications for the system's dynamics, for the process taking place on it, and ultimately for its function. For instance, the choice of a particular class of connectivity metrics induces a corresponding change in combinatorial, topological and geometrical properties of the associated network structure and phase space geometry and, therefore, different phenomenology and physics (del Genio et al., 2016).

The following questions are addressed: how can network structure be modified to incorporate network detail? What may be the phenomenological consequences of such changes in network structure? Can we account for more neural phenomenology by changing network structure? Is there experimental evidence for a given generalised structure? How robust is the system's behaviour with respect to changes in its basic structure? Assuming a simple, undirected, unweighted and static network structure as the ground-level, there are essentially three ways in which unaccounted for neural properties can be addressed: 1) considering different properties of the original structure, e.g. properties of the connectivity matrix; 2) understanding the system as a network structure with different allowed constituent properties; 3) understanding the system as a qualitatively different network-based structure.

### 4.1. Roads less travelled in standard network neuroscience

*4.1.1. Degrees of freedom*

The general neuroscience problem of defining relevant neural units and relevant degrees of freedom is replicated when equipping the system with a network structure. At a network level, the discretisation process may in principle be predicated upon various properties, the identification of truly functional constituent units in real space being only one of them.

In real space representations, the system's degrees of freedom are most often identified with nodes, irrespective of the scale at which a network structure is defined, but particularly at system level. On the other hand, in statistical mechanics, the system's degrees of freedom are identified with links, whereas the number of particles play the role of volume in classical physical systems (Gabrielli et al., 2019). In the corresponding *dual networks*, nodes are turned into links and, conversely, links become nodes (Presigny et al., 2024). While these two networks are in some sense equivalent, a link-based approach may for instance allow defining fine-grained vasculature data at all length scales and therefore also measuring blood flow conductance, current and inferring pressure differences for each link (Di Giovanna et al., 2018; Kirst et al., 2020).

Another important aspect of network structure is that, however defined, the degrees of freedom can have their own spatio-temporal dynamics in real space (Korhonen et al., 2021). Thus, at the level of dynamics and function, it may be appropriate to think of the system as a fluid structure where both nodes and links may be non-stationary (Solé et al., 2019). Nodes may appear or disappear, merge as a result of physiological or pathological conditions at various spatial and temporal scales or change their spatial location. For example, at subneural scales, spine motility (Bonhoeffer and Yuste, 2002) may be thought of in terms of moving nodes. Nodes may also constitute local subspaces in the codomain onto which they are projected but may result from non-local subspaces of the domain space. Similarly, the activity of neurons spiking at the same time can be identified with the nodes of a network whose links are the neurons themselves (Curto and Itskov, 2008; Morone et al., 2019, 2020).

*4.1.2. Directionality and reciprocity*

One property characterising neural activity seldom included but that can readily be accounted for in a standard simple network structure is flow directionality. Directionality may characterise both real and phase space neural activity. In the latter case, brain activity is thought of as discrete dynamical system, whose trajectories form a directed network in state space, wherein each node, representing a state, is the source of a link pointing to its dynamical successor. Directed networks qualitatively differ from their undirected counterparts in the system's combinatorics and statistical mechanics (Boguñá and Serrano, 2025), but also in important aspects of the dynamical processes such as synchronisation (Muolo et al., 2024), pattern formation (Asllani et al., 2014), phase-transitions (Fruchart et al., 2021). The presence of asymmetric connectivity is associated with the emergence of some important features: on the one hand, spontaneous activity is characterised by time scales and corresponding oscillatory modes different with respect to those emerging from symmetric connectivity (Chen and Bialek, 2024). On the other hand, when perturbed, systems with asymmetric connectivity undergo complex transients, with time scales induced by different aspects of the connectivity matrix with respect to those of symmetric connectivity (Grela, 2017). Moreover, in the presence of asymmetric interactions, fluctuations can get locally enhanced before propagating through the system promoting collective qualitatively changes in large scale collective behaviour in globally ordered systems (Cavagna et al., 2017). This can be explained in terms of frustration, which arises when competing interactions prevent the system from finding a configuration that minimises the total energy leading to a complex disordered state (Vannimenus and Toulouse, 1977). Frustrated closed-loop motifs disrupt synchronous dynamics, allowing the coexistence of multiple metastable configurations (Gollo and Breakspear, 2014; Saberi et al., 2022).

More generally, directed links metrics induce different physics. While symmetric connectivity readily accounts for equilibrium systems, asymmetric coupling matrices are associated with open out-of-equilibrium systems, where detailed balance is broken (Nartallo-Kaluarachchi et al., 2024). Such systems exchange energy with the external environment, allowing effects such as gain and loss, and *non-reciprocity* (Bowick et al., 2022). In many-body systems, non-reciprocity leads to the dynamical recovery of spontaneously broken continuous symmetries (Fruchart et al., 2021). Conversely, non-reciprocal coupling per se usually implies non-zero energy and information flows (Loos and Klapp, 2020). This has not only theoretical but also methodological implications. For instances, choices associated with links and the way these are constructed, e.g. hybrid reconstruction with space- and time-varying properties, represent not only a technical but also a theoretical challenge, in that they induce spaces with non-trivial geometries and corresponding physics.

*4.1.3. Beyond network topology*

Brain modelling typically focuses on combinatorial and topological properties, neglecting neural networks' physical aspects. The underlying space is treated as a topological space, i.e. a set of objects equipped with a set of neighborhood relations and no usual metric needs to be defined. It is therefore effectively treated as a non-metric space. In such models, nodes and links are treated as dimensionless entities. However, the physical size of neural material affects network geometry at all scales. The fact that physical wires cannot cross imposes limitations on the system's structure (Bernal and Mason, 1960). In particular, the path chosen by neural fibres may be characterised by tortuosity as a function of node and link size and density (Dehmamy et al., 2018). Thus, the system's structure is ultimately determined not only by operators associated with the connectivity matrix, but also by the network's 3D layout (Cohen and Havlin, 2010). For a given network adjacency matrix, there is an infinite number of configurations differing in node positions and wiring geometry, those of which that can be bijectively mapped into one another through continuous bending, and without link crossings forming *isotopy classes* (Liu et al., 2021). The geometry of connectivity may have an important impact on cortical dynamics and function (Knoblauch et al., 2016). For instance, lobal brain activity patterns may result from excitations of brain geometry's resonant modes, which may better capture important properties of spontaneous and stimulus-induced activity with respect to connectivity-based models disregarding neural surface's geometry (Pang et al., 2023). Thus, methods may be needed which can distinguish between topologically equivalent manifolds with different geometries (Chaudhuri et al., 2019).

The no-crossing condition also affects the system's mechanical properties. While at low densities the system displays a solid-like response to stress, for high densities it behaves in a gel-like fashion (Dehmamy et al., 2018). The brain's mechanical properties play a critical role in modulating brain anatomy, dynamics and ultimately function (Goriely et al., 2015). Due to its softness, brain tissue displays a range of mechanical features: it is essentially elastic for small deformations (Chatelin et al., 2010), but inelastic and deformation rate- and time-scale-dependent for large ones (Fallenstein et al., 1969).

Overall, numerous questions are still to be fully addressed: what's the relationship between topology and geometry in anatomical networks? In particular, to what extent does topology determines geometry? How do the geometric constraints on wiring affect brain structure, dynamics, development, evolution, functional efficiency and robustness to various sources of damage?

*4.1.4. Learning rules and adaptative networks*

One way in which neural populations adapt to environmental challenges is by changing their configuration. At time scales longer than those of sensory-motor processes, this typically involves plasticity mechanisms. At the algorithmic level, homeostatic plasticity mechanisms constitute slow negative feedback loops (Zierenberg et al., 2018). Various studies incorporated simple plasticity mechanisms into large-scale network models, showing that this may affect network topology (Avalos-Gaytán et al., 2012, 2018) and give rise to rich dynamical phenomena including intermittency (Skardal et al., 2014), multistability (Chandrasekar et al., 2014) and criticality (Magnasco et al., 2009) or explosive synchronisation (Avalos-Gaytán et al., 2018).

From a network viewpoint, various questions are still unresolved: on what network aspect (including at what network scale) does plasticity operate? What algorithmic properties do plasticity mechanisms possess? Is the system adaptive, i.e. are there feedback mechanisms connecting topology to dynamics? If so, how do they affect function?

**4.2. Beyond ground-level network structure**

If the simple network structure fails to incorporate essential properties of neural anatomy and dynamics, its modelling power should be addressed by allowing structures with different properties and appreciating the changes that these may produce. Insofar as the simple network structure can be thought of as a ground level of network structure, new network classes can be obtained by relaxing some of its properties.

*4.2.1. Recurrency and feedback loops*

A set of anatomical properties of neural circuits generally not incorporated in standard system-level network representations is represented by recurrency. In terms of network properties this translates into *self-loops*, i.e. links connecting a node to itself, and *cycles*, i.e. closed paths with the same starting and ending node (Douglas and Martin, 2007; Fan et al., 2021). Recurrent interactions play a major role in dynamics, leading to chaotic dynamics (van Vreeswijk and Sompolinsky, 1996; Pernice et al., 2011, 2013). Moreover, feedback loops are an essential ingredient in both dynamics and computation (Alon, 2007; Zañudo et al., 2013). For instance, multistability and sustained oscillations respectively require positive and negative feedback loops (Thomas, 1981). Moreover, various neuromorphic devices (Marković et al., 2020) with recurrent connectivity, including liquid and solid state machines, echo-state networks, and general deep neural networks (Maass et al., 2002), and physical reservoir computing devices exploiting physical systems dynamics as devices (Tanaka et al., 2019) display information-processing capabilities. In such devices, multiple recurrently connected dynamical systems are used to implement nonlinear mappings of input signals into a high-dimensional state space using.

*4.2.2. Higher-order network structure*

Perhaps the most natural generalisation of network structure consists in changing its combinatorial properties by relaxing the pairwise (*dyadic*) character of connectivity (Lambiotte et al., 2019; Battiston et al., 2020, 2021; Bianconi, 2021; Bick et al., 2023). This may be done in various ways (*See* A4). While in standard networks interactions are associated with links connecting two nodes, graphs can be generalised to include *hyperlinks*, i.e. links connecting more than two nodes (Ghoshal et al., 2009). Interactions could also in principle involve structures of different orders. *Simplicial complexes* allow defining interactions across orders (nodes, hyperlinks or simplices) (Giusti et al., 2016). In simplicial complexes, state variables used to describe the dynamical system can be associated with structure at any order. Thus, for instance, the state of a link can influence not only the state of its associated nodes, but also that of the higher-order interaction structures it belongs to, and such a system's overall dynamics ultimately results from the time-varying interactions across all orders. Moreover, a node may regulate the interaction between two other nodes, either facilitating or inhibiting it (Sun et al., 2023; Niedostatek et al., 2024).

Higher-order topology has an important role in determining both dynamical and functional network properties. For instance, the dynamical ordered state has minima corresponding to single *homology classes* of the simplicial complex (Millán et al., 2020). Furthermore, coupled oscillator networks have fixed points consisting of two clusters of oscillators that become entrained at opposite phases and which can be thought of as configurations with information storage ability. Topology determines the small subset of the fixed points which are stable (Skardal and Arenas, 2020).

What experimental evidence is there for or against the existence of such a structure in neural systems? The structural aspects of higher-level interactions in the network structure of brain dynamics have long been addressed. Early studies suggested that real space neural activity may almost completely be explained in terms of pairwise correlations (Schneidman et al., 2006; Merchan and Nemenman, 2016). However, this experimental result could crucially hinge on the overall size of the considered cell population, and higher-level correlations may be necessary to account for larger populations' neural activity (Yeh et al., 2010; Ganmor et al., 2011; Giusti et al., 2015; Reimann et al., 2017). Experimental evidence suggests that dynamical correlations between pairs of neurons are more significant when these belong to higher dimensional structure (Reimann et al., 2017), although recent results suggest that brain activity is dominated by pairwise interactions (Huang et al., 2017; Chung et al., 2025). In phase space, a higher-level structure may be induced by the intersection of place fields of neurons firing within the same theta frequencies cycle. Under certain conditions of the place fields, the homology of the simplicial complex induced by the intersecions is equal to that of the underlying space, so that this structure effectively constitutes a faithful internal representation of the stimulus space ignoring finer phase-modulated spike timing effects (Curto and Itskov, 2008). Place field intersections also induce a metric providing relative distances between cell groups. This yields a faithful geometric representations of the external physical space somehow independent of the specific nature of the place fields.

On the other hand, the interactions between structure of different dimensions in principle afforded by a truly simplicial structure, have not been investigated in earnest yet. In a spatially embedded physiological context, this would almost necessarily involve cross-talk between dynamics at different spatial but also temporal scales.

Supposing that neural systems indeed present significant non-dyadic structure, for instance that higher-order dynamical systems do not result from some coordinate transformation of dyadic network dynamical systems (Bick et al., 2023), what is the neurophysiological meaning of such a class of structures? Can computation be performed in such structures? If so, which ones? How is it implemented by neural systems?

But what does this structure say about the mechanisms underlying its emergence and generating observed phenomenology? On the one hand, it has been pointed out that high order structure of the system's emergent properties does not necessarily require high-order terms in the underlying dynamical law or in the Hamiltonian, and that even high-order methods relying on pairwise statistics (e.g. simplicial complexes built from a correlation matrix) may miss significant information only present in the joint probability distribution but not the pairwise marginals (Rosas et al., 2022; Robiglio et al., 2025). On the other hand, observed phenomena are not always a good proxy for the underlying generating mechanisms. In particular, the presence of statistical synergy does not imply genuinely non-decomposable interactions per se, as observable patterns may emerge from additive dynamics and pairwise interaction sequences, and even if complex collective behaviour can in principle involve irreducibility it often does not (Ji et al., 2023).

Structural descriptions based on higher-level generalisations face the standard problem in network modelling, i.e. mapping the network structure on appropriate aspects of the system, but, in spite of the restriction on the admissible contiguity laws, have an otherwise rather intuitive meaning, both in real and in phase space. On the other hand, dynamical descriptions are more problematic. For instance, while it is reasonable to assume that neural computation resorts to some form of discrete calculus and that it may integrate information across scales, it is not straightforward to understand neural dynamics and function in terms of standard exterior calculus and co-boundary operators. Furthermore, it is not clear to what extent brain dynamics presents meaningful interactions across structures of different dimensions.

*4.2.3. Generalised interaction types*

A further network structure generalisation consists in allowing multiple types of interactions between nodes (De Domenico et al., 2013; Boccaletti et al., 2014, 2023; Kivelä et al., 2014; Bianconi, 2018). In this class of structures, nodes may exist at different layers, with a connectivity structure in principle independent at each layer. Intra-layer links belong to the same layer and inter-layer links connect the projections of the nodes at different layers (*See* A5). The layers of a multiplex network can account for different interaction phenomena such as information transfer or the ability to synchronise (De Domenico, 2017; Buldú and Porter, 2018). Moreover, at least *prima facie*, this class of structures appears as a natural representation of interdependencies among different systems (both within and without the brain) and can therefore be used to assess properties such as stability (Bonamassa et al., 2021), robustness and vulnerability (Buldyrev et al., 2010; Gao et al., 2011; De Domenico et al., 2014), or to understand the nature of interactions, e.g. competition (Danziger et al., 2019). Such a structure can highlight the role of connectivity, particularly of connector nodes in the modulation of bare dynamics or of processes unfolding on the network (Aguirre et al., 2013, 2014; Buldú et al., 2016). Not only does the interaction of a given subgraph with other nodes in the network affect whether that subgraph corresponds to a fixed-point support (Morrison and Curto, 2019), but the type of node (peripheral or central) acting as connector between subnetworks affects dynamics and processes in each of them (Aguirre et al., 2013, 2014). Note that this construct is not different from that of a standard network but rather a change in the way the relation matrix is segmented.

Unpacking information that may be hidden in standard collapsed representations (Cardillo et al., 2013; Zanin, 2015; Papo and Buldú, 2018) may better account for interdependencies of interacting units within single network units and may reveal structural and dynamical properties of biological networks, for instance synchronisation properties which may be opposite to those operating in isolated networks (Aguirre et al., 2014).

Multilayer networks may naturally account for the layered structure of cerebral and cerebellar cortices (Huber et al., 2021) but also for interactions between neural populations in the cerebral cortex and separable subsystems such as the neuromodulatory system (Brezina and Weiss, 1997; Brezina, 2010), as well as the relationship between neural and extrinsic systems, e.g. the heart or the breathing system, as in a network physiology approach (Bashan et al., 2012). A multilayer (and multiplex) interaction structure has also been associated with the interactions between brain regions at different frequency bands, each band corresponding to a different layer of a multiplex/multilayer network undetected when averaging activity across layers (De Domenico, 2017; Buldú and Porter, 2018). Furthermore, various results point to the possibility of using multilayer brain networks as biomarkers of brain degenerative diseases such as Parkinson's disease, mild cognitive impairment of Alzheimer's disease (De Domenico et al., 2016; Echegoyen et al., 2021).

A subclass of multilayer networks is represented by temporal networks, wherein each layer corresponds to the structure at a particular time step, and layers are connected through unidirectional time-ordered links (Holme and Saramäki, 2012). In temporal networks, nodes are related to each other via causal or *time-respecting paths* (Holme, 2015), and dynamic interactions' complex temporal structure may lead to history-dependent paths with long-term memory (Scholtes et al., 2014). Higher-order dependencies between nodes imply that causal paths can be more complex than those induced by static and aggregated networks and can affect topological network properties, e.g. node centrality (Scholtes et al., 2016), or community structure (Rosvall et al., 2014; Peixoto and Rosvall, 2017), dynamical processes, e.g. diffusion and dynamical processes (Ghosh et al., 2022) and the controllability (Zhang et al., 2021; Li et al., 2017).

At long time scales, brain fluctuations are characterised by non-trivial dynamical and statistical properties such as intermittency, scale invariance and long-range temporal correlations (Novikov et al., 1997; Allegrini et al., 2010; Fraiman and Chialvo, 2012; Papo, 2014a). Multilayer temporal networks may capture non-trivial higher-order cross-order interactions, including cross-memory among neural populations, with complex fluctuating dynamics and nucleation or coalescence of neuronal populations (Gallo et al., 2024). However, whether such a symmetry breaking is present in brain activity and its functional meaning is still poorly understood.

## 4.3. Beyond single networks

Relaxing simple network properties gives rise to generalised possibly associated with profoundly different phenomenology but in some sense similar networks. However, the brain could be endowed with a structure that does not stem from property relaxation, and that may be qualitatively different from that of a simple network, ultimately changing the very essence of brain networkness.

### 4.3.1. From single networks to network ensembles and sequences

In essence, most network models of brain activity constitute field theories studying the time evolution of relevant variables measured at each point in time and on a finite number of points in space (Mikaberidze et al., 2025). Insofar as the relevant field variables are inherently fluctuating quantities, it is natural to describe the probability of field states in terms of ensembles incorporating the uncertainty about the system's state or, equivalently, describing the system's possible states and their structure.

Rather than focussing on the relational structure to learn about the topological and geometrical network properties of neural systems, a useful representation may highlight the statistical properties of relations. Networked structures are then described by statistical models that specify a probability distribution over a set of graphs, e.g. a probability of observing a given set of relations (Dichio and De Vico Fallani, 2022) and the quantities of interest are the set of properties of such spaces (Kahle, 2014) (*See* A6). The frequency with which topological properties appear and their significance are explained in terms of probability distributions. Thus, the system is characterised not only in terms of topological invariants but also of their scaling properties, e.g. with system size or dimension. This framework's dynamical counterpart is represented by the *path-integral approach*, where the system's dynamics is represented by weighted sums of all possible paths the system can take. In a conceptually similar approach, each node can be understood as a superposition of multiple states (Ghavasieh and De Domenico, 2022).

The shift between single network to network ensembles highlights various aspects corresponding to different cuts into the relevant space. First, the relevant structure is not that of single realisations of a process (or of averaged or steady-state equivalents) or of a specific scale or scale range. These structures induce an effective thermodynamics, whose thermodynamic potentials and their non-analytical points identify corresponding phase transitions (Meshulam and Bialek, 2024). Second, proper brain structure and function representations may contain a relationship between these representations. This can be thought of in various ways, e.g. in terms of the minimum and maximum coupling levels, which act as energy levels in Hamiltonian systems (Santos et al., 2019), below and above which topological invariants vanish (Santos et al., 2009) or as the limit of a sequence of graphs, e.g. a *graphon* (Lovász and Szegedy, 2006) and effectively treated as a dynamical system (Bick and Sclosa, 2024).

### 4.3.2. Models of network models

A more fundamental way to understand relationships across scales consists in conceiving of the network structure as an *effective field theory* of brain structure and dynamics, i.e. a description of a system's physics at a given scale up to a certain level of accuracy, using a finite number of variables that parametrise unspecified information in a useful way (Georgi, 1993) (*See* A7). Indeed, in both anatomical and dynamical brain network representations, nodes and links, which constitute the microscopic scale of a network representation at a given scale, can always be understood as resulting from renormalisation of neurophysiological properties at lower scales and each degree of freedom effectively constitutes a kinetic model of phenomena at lower scales. Particularly at meso- and macroscopic neural scales, each node is then an asymptotically stable invariant subset of the phase space (in the simplest case a fixed point) in a renormalisation flow across scales.

Renormalisation theory shifts the focus from the investigation of the outcomes of a given model to the analysis of models themselves, by relating models of the same system at different scales along a renormalisation trajectory or grouping models of different systems sharing the same critical behavior and exhibiting the same large scale behaviour (*See* A7). The renormalisation framework highlights scale-dependence of interactions in a systematic way and allows investigating the level at which non-random structure emerges, the relationship of such structure with the one present at other levels and ultimately the possible ways in which spatio-temporal patterns are converted into macroscopic dynamics and function.

Network sequences induce corresponding spaces with rich non-random structure. At each renormalisation flow stage, one may consider the coarse-grained structure emerging above the level of individual nodes in the system's hierarchical organisation, whose nodes correspond in some sense to communities, and whose links represent members shared by two communities (Pollner et al., 2006). The presence of structure at various scales (not all of which ought to be functionally meaningful) reflects a property of biological systems, which may present different out-of-equilibrium properties at different scales, or equilibrium properties at certain scales but not at others (Cugliandolo et al., 1997; Egolf, 2000).

The renormalisation flow can be understood as dynamics on the space of field models, and it is important to understand the extent to which it operates in a *functorial*, structure-preserving manner, linking different field models and their properties, i.e. how it preserves properties not only of its space components but also of maps between them (Ghrist, 2014) (*See* A8).

Note that brain network renormalisation typically dispenses with the treatment of infinities involved in the transition between essentially continuous anatomical or dynamical fields, and network structure. This mapping is dealt with through discretionary steps the consequences of which are poorly understood (Korhonen et al., 2021).

*4.3.3. Emergence of network structure and function*

It is straightforward to understand network ensembles and sequences in terms of structure emergence. The structure's statistical properties emerge naturally from constrained entropy maximisation, each constraint giving rise to different models (Radicchi et al., 2020) a reasonable model at long, e.g. evolutionary time scales (*See* A9). In this vein, for example, mean-field representations can be thought of as maximum entropy models for the topology of direct interactions, whereas network models as that of paths (Lambiotte et al., 2019). More generally, a system's organising principles can be thought of as the result of underlying non-equilibrium growth and development mechanisms (Betzel and Bassett, 2017).

Renormalisation, perhaps the most general conceptual representation of emergence at any scale, is in general understood as an analytical tool to highlight neural structure (*See* A9). In this context, coarse-graining has two contrasting effects: on the one hand, it is necessarily associated with information loss. On the other hand, it reduces noise and increases the strength of relationships, so that structure may emerge far from the micro-scale, where macro-states have stronger dependencies (Hoel et al., 2013). Emergent behaviour can be transient, context-dependent and non-local in real space or time (Varley, 2022), and behaviour at one scale may not be well predicted by behaviour at a finer scale (Wolpert and MacReady, 2000). Morevoer, causality may be circular, with large scale behaviour dictating the one at lower scales (Haken, 2006b) (See A9). The following important questions are often addressed: what makes a neuronal unit reducible to a node? Can coarse-graining allow neglecting hardware heterogeneity, e.g. glial cells? What structure does the renormalisation flow preserve? To what extent is functionally relevant information retained or lost in coarse-graining?

Renormalisation can also be thought of as a genuinely functional neural process. In this sense, network structure emergence can be distinguished from the emergence of function. Function emerges from one particular coarse-graining procedure (which may not necessarily correspond to real space renormalisation) (Bradde and Bialek, 2017). For instance, in the sensory domain, network structure constitutes the *nerve covering* induced by boundary conditions emerging from dynamical annealed disorder associated with neuronal populations' receptor fields (Curto, 2017). In this framework, nodes and links are emergent properties, rather than a structure *a priori* (See A9). Likewise, geometry can be seen as an emerging property of single neurons' physiology and of the functional architecture through which these local properties are renormalised. Whether the emerging structure is fundamental or a manifestation of a more primitive, *pre-geometric* reality (Bianconi et al., 2015) depends on whether it has functional value or not.

The corresponding questions are: how does behaviour emerge from its spatio-temporal dynamics? If renormalisation represents how function emerges, to what extent do appropriate representations depend on of the specific renormalisation process?

## 5. Pathways for network neuroscience

In addressing possible future avenues for a network understanding of the brain great emphasis is typically put on improving experimental techniques, for instance, electron microscopy reconstruction is expected to significantly improve accuracy and scope with respect to traditional electrophysiological techniques and may help constraining computational models (Litwin-Kumar and Turaga, 2019) and on mathematical and physical modeling and in data analysis techniques (Goodfellow et al., 2022). However, advances may come from better knowledge concerning fundamental aspects of brain functioning and, from changes in some conceptual aspects of the network-brain association at the heart of network neuroscience.

Here we discuss three main conceptual axes along which network neuroscience may evolve: (i) the use of function to gauge brain models; (ii) how network theory should help in advancing neuroscientific knowledge and conceptual apparatus; (iii) how phenomenology is explained.

### 5.1. Gauging structure through function

One of the most fundamental endeavours of network neuroscience is to understand how network structure is related to brain function (Ito et al., 2020). Thus, network structure can be associated with some fitness for specific tasks. While not all observed structure has functional meaning, and not all observed features optimise function but may instead be a byproduct of the way the network evolved (Solé and Valverde, 2020), without a proper theory of its function it is in general arduous to explain observed anatomical and dynamical structure and generative models are underdetermined both at experimental and at longer time scales (Doyle and Csete, 2011). Brain networks and the dynamical processes occurring on them are to a large extent the result of evolutionary, learning and adaptation processes, through which the brain solves computational problems necessary for survival, which in turn arbitrate trade-offs among available resources. Classical statistical physics approaches do not incorporate the notion of function, partly due to the fact that large non-biological disordered systems such as glasses do not arise through evolutionary processes (Advani et al., 2013).

The relationship between structural properties, e.g. topology, and function may suggest features essential to appropriate phenomenological models. For instance, if function and functional dynamics are respectively associated with some structural universality class and *topological phase transitions*, i.e. qualitative changes in topology, then this should be accounted for and a corresponding physiological mechanism should be found. While altering the microscopic scale affects the resulting physics, the question is not only whether the associated phenomenology constitutes a good descriptor of brain dynamics and function but also whether there are elements suggesting its plausibility.

The relationship between brain structure, dynamics and function is in many ways a complex one. First, the properties of a networked dynamical system do not trivially stem from either local dynamics or network structure alone, but from the interaction of the two (Curto and Morrison, 2019). Second, while deterministic macroscopic order can govern function, e.g. learning, such structure can arise in ways that are independent of the details of network heterogeneity (Advani et al., 2013). Third, while tens of neurons may be sufficient to identify the network's dominant variability modes (Williamson et al., 2016), a given network's function can vary in a context-dependent way (Biswas and Fitzgerald, 2022), although different strucutres tend to be optimal for different tasks. For instance, information flow and response diversity are optimised by different circuits (Ghavasieh and De Domenico, 2024). Conversely, different networks can give rise to similar function. Overall, how network structure contributes to neural networks' dynamical and functional properties such as *sloppiness* and *degeneracy* (Gutenkunst et al., 2007; Machta et al., 2013) is still poorly understood (*See* A10). Fourth, structural complexity does not necessarily lead to functional complexity, e.g. to heterogeneous responses to perturbations. Functional heterogeneity is in general a genuine emergent property which cannot be deduced from the system's structural properties e.g. structural heterogeneity or dynamical properties (Ghavasieh and De Domenico, 2022), and which can give rise to rather non-trivial phase space configurations (Stadler and Stadler., 2006) (*See* A10). Ultimately, which aspects and properties of network structure are necessary in a brain model and, as a result, which methods should be summoned to represent them (Giusti et al., 2015; Curto, 2017) depend on the properties of such mapping.

Finally, while a network's function can help understanding both its structure and complex dynamics (Chklovskii et al., 2004; Lau et al., 2007), function itself may not always be obvious a priori and may have a non-trivial relationship with bare dynamics (Papo, 2019a). The general dearth of neural, particularly functional stylised facts induces a circularity of functional brain networks: the incomplete knowledge of the algorithmic and implementation aspects of neural computation, even at single neuron scales (Moore et al., 2024), biases the segmentation at microscopic scales, giving rise to network structure that is not necessarily functionally meaningful. sometimes, models contain mechanisms which incorporate properties which appear plausible. For instance, network structure dynamics could be understood as emerging from a non-equilibrium dynamics similar to that of the *network geometry with flavour* growth model, where the flavour parameter may appear a good candidate model for neuromodulation (Bianconi and Rahmede, 2015; Bianconi et al., 2016). Often models also aim at replicating some of the system's ostensible generic statistical or dynamical properties, e.g. its scaling of fluctuations. However, often these can arise in rather different ways (Morrell et al., 2021), and their functional properties are in general not directly tested but only inferred based on prior knowledge.

*5.1.1. Universality*

In some sense, understanding how robust network structure is with respect to both biological detail and network specification is a question germaine to the issues of neural functional equivalence and switching. Indeed, universality constitutes a form of robustness (Lesne, 2008b). Moreover, from a functional viewpoint, an important question is the extent to which function is robust to changes in structure. A nested question is related to the scale-dependence of such relationship, i.e. the scales at which the structure-function map induces qualitative changes.

A fundamental issue in the determination of network structure's role in the brain is the extent to which dynamical emergence is a property of network structure, e.g. topology, independently of the specific properties of node dynamics. On the one hand, emergent dynamics is not necessarily inherited from intrinsically oscillating nodes or induced by the characteristics of forcing stimuli but may arise from the coupling structure (Morrison et al., 2024). This is for instance the case of threshold-linear networks (Morrison and Curto, 2024). On the other hand, empirically observed fluctuation scaling properties can be achieved by imposing specific nodal properties, e.g. a particular type of neuron excitability (Buendía et al., 2021).This could for instance be implemented by neural apparatus in which the global coupling strength would be normalised by the average coupling strength per node, so that the dynamics would be invariant under scaling of the adjacency matrix, sterilising the role of the network from the specific properties of the nodes (Nishikawa and Motter, 2016).

In a statistical physics sense, universality reflects the fact that many systems possibly differing in their microscopic properties, can nonetheless be classified into a small number of *universality classes* defined by their *scaling exponents*, which quantify a system's relationship between different scales (*See* A7). Universal relations arise when the changes caused by modification of microscopic parameters are effectively summarised by a small number of phenomenological parameters (Goldenfeld, 1989). Complex systems such as the brain may exhibit instability of renormalisation, i.e. may fail to converge to a stable fixed point, within a *topological class*, comprising systems or states sharing the same fundamental topological properties, even for stationary combinatorics (Martens and Winckler, 2016).

While the temporal structure of avalanches shows signs of universality (Friedman et al., 2012), one important question is that of determining how network properties may contribute to its emergence. In correlated inhomogeneous structures, universal behaviour is comparable to the one characterising continuous field theories of system with non-integer dimension, and the relevant control parameter for universal behaviour on inhomogeneous structures is the spectral dimension (Millán et al., 2021a).

*5.2. A neuro-inspired network science*

What does network structure tell us about fundamental properties of brain dynamics and function? Can we express how efficiently the brain carries out its functions or how it can withstand environmental challenges, possibly changing as a result of them, in terms of network structure?

For many systems it is natural to relate properties such as *robustness* and *efficiency* to the topological properties of its network structure (Ma et al., 2009; Estrada et al., 2012; Faci-Lázaro et al., 2022). However, there is no guarantee that the way efficiency and robustness (or, equivalently, resilience and vulnerability) are usually defined (Kitano, 2002b, 2007; Lesne, 2008b; Liu et al., 2022; Schwarze et al., 2024) is actually a good indicator of functional robustness (Papo and Buldú, 2025a). Furthermore, there is little knowledge of the topological properties that may covary with functional robustness and of the relationship between robustness, degeneracy and evolvability in the brain (Wagner, 2008; Masel and Trotter, 2010; Whitacre and Bender, 2010; Whitacre, 2012). Future research should quantify properties such as robustness and efficiency in a way that is functionally meaningful (Levit-Binnun and Golland, 2012; Papo and Buldú, 2025a). This will imply a conceptual effort and perhaps the adoption of meaningful metaphors.

*5.3. From causal to topological explanations*

A fundamental question not often addressed is whether network properties can be used to explain neural function.

*Causal explanations* are thought of as essential to the scientific method (Livneh, 2023). Causal explanations account for observed process or performed function in terms of chains of causal factors or interactions bound by spatio-temporal continuity and statistical relevance (Van Fraassen, 1977). However, in systems with a great number of non-linearly and non-locally interacting units such as the brain, causal chains may be difficult both to observe and to define, as global parameters emerging from the intrinsic interactions among the individual parts of the system may in turn govern their behaviour (Haken, 2006a,b). Alternative types of explanation are often thought of as satisfying accounts of observed phenomena. For instance, mechanistic explanations aim at highlighting neurophysiological *mechanisms* i.e. "entities and activities organised in such a way that they are responsible for the phenomenon" (Illari and Williamson, 2012).

Results from research fields ranging from condensed matter physics (Thouless et al., 1982; Bowick and Giomi, 2009), to quantum computing (Collins, 2006) and data mining (Rasetti and Merelli, 2015) indicate that complex system's phenomenology can be explained in topological terms (Huneman, 2010; Kostić, 2016; Tozzi and Papo, 2020). Topological descriptions may help establishing under what conditions network structure, and under which conditions network structure is relevant, and predicting and acting upon brain activity (Papo, 2019b). While various factors, including function and energetics may account for structure and dynamics, a still rather poorly understood question is the extent to which structure, particularly topology, explains or constrains function, or rather constitutes a mere by-product of it.

## 6. Concluding remarks

We discussed on the one hand neural structure typically drastically simplified in standard network neuroscience models of brain anatomy and dynamics and on the other hand possible alternative network structure classes and the potential benefits that these may offer in terms of ability to account for known neural phenomenology or to reveal as yet unknown one, distinguishing between computing different properties of a standard network structure and changing the structure itself.

In essence we addressed two dual questions: to what extent does adding biological detail qualitatively change network models of brain activity? How universal is network structure? Appropriate phenomenological descriptions of a system always contain a universal part and a few detail-sensitive constants (Goldenfeld et al., 1989). The quest for the appropriate level of detail characterises the study of most complex biological systems. In some sense, understanding the brain as a networked system boils down to determining whether a statistical mechanics approach makes sense and at which scales details matter (Cavagna et al., 2018). In network neuroscience, it is important to understand to what extent network structure not explicitly incorporating the important neural properties mentioned hitherto nonetheless recovers good approximations of brain structure, dynamics and ultimately function.

What can generalised structures change with respect to standard, network models? Over and above different phenomenology, generalising network structure can provide with different ways to conceive of brain anatomy, dynamics and function and, more fundamentally, to explain neurophysiological phenomena. However, generalised structures face the same fundamental issues related to intrinsicality, universality (intended as robustness to changes in neurophysiological detail), and functional meaningfulness of standard network models: is structure intrinsic? If so, how does it allow the system to carry out the functions it is assigned? What aspect of the neural system's network structure is functionally meaningful? To what extent is such a structure universal? How can we decide whether a given structure is a mere extrinsic description or can be thought of as part of its intrinsic *modus operandi*? While whether there exists an appropriate structure representing a given dynamical system may be a question of context (Bick et al., 2023). Answering these fundamental questions will require incorporating function but also a better characterisation of neurophysiological stylised facts and of the structure-dynamics-function relationship.

Finally, throughout, we mainly discussed the extent to which a network representation reflects the way the system may work, rather than how such a structure allows investigating it theoretically or experimentally. The method used to investigate a system (e.g., the process used to explore a network) and the functions that the system actually implements are somehow intertwined and often equated, and so are a given structure's information content and the dynamical aspects that this structure supports. For instance, a given space parametrisation may be expedient in a particular context, but may not reflect the sytem's underlying functional geometry, affording an *extrinsic* embedding-dependent view of the true underlying space (Pennec et al., 2006; Lenglet et al., 2006). In the space underlying a given representation, the allowed operations may also not reflect the computations performed by the system. Likewise, while a given structure may be associated with a certain amount of information, that doesn't entail that such information is actually transferre or computed or that is functionally relevant.

## Appendices

### A1. Network and network properties

- In its most general form, a network is a structure $\mathcal{N} = (V, E)$, where $V$ is a finite set of *nodes* or *vertices* and $E \subseteq V \otimes V$ a set of pairs of *links* or *edges* $V$. All the information in a network structure is encoded in a connectivity matrix, which can take various forms, e.g. the *adjacency matrix* $A \in \mathbb{R}^{N \times N}$ whose entries $a_{ij}$ are equal to 1 if nodes $i$ and $j$ are adjacent, and equal to 0 otherwise; or the *Laplacian matrix* $\mathcal{L} = \delta_{ij} \sum_k (A_{ik} - A_{ij})$ where $\delta$ is the Kronecker function.
- The links can carry a *weight*, parametrising the strength of interactions, giving rise to a structure $\mathcal{N} = (V, E, w)$, where $w$ is a real (or complex) function $w: E \to \mathbb{R}$ (or $\mathbb{C}$), and a *direction*, in which case $E$ comprises ordered pairs.
- *Combinatorial properties* are properties related to counting and structural properties of the graph itself, such as the number of vertices, edges, or cycles. Combinatorial graph theory focuses on the exact number and arrangement of nodes and links.
- *Topological properties* are properties that are invariant under continuous deformations, such as stretching, bending, or twisting, without tearing or gluing. Two space are said to be *topologically equivalent* if they can be continuously deformed into one another. Topological properties include connectedness or the *genus*, intuitively counting the number of holes or handles in a surface.
- *Geometric properties* concern the physical arrangement and characteristics of the graph's elements in space. Geometric properties include link length and the angles between them, as well as the graph's shape.

### A2. Neural network modelling

A typical firing-rate neural network model describes the time evolution of $N$ recurrently connected nodes $x_i$

$$\dot{x}_i = F\left(x_i, \sum_{j \neq i}^N A_{ij} H_i(x), \beta_i b(t)\right) \quad [1]$$

where $\dot{x}_i(t)$ denotes the *i*th node's state variation rate, $F(x_i, 0)$ the dynamics of the isolated node, $A_{ij}$ the connectivity matrix, $H_i$ the drive from other nodes on the *i*th node, and $\beta_i$ quantifies the way a time-dependent input signal $b(t)$ affects node $i$.

Overall, the system's collective dynamics depends on each node's intrinsic dynamics F and on the graph structure, encoded in the connectivity matrix $A$ and the coupling function $G$. Thus, both local and global dynamical properties are influenced by the interaction properties e.g. the graph spectrum is related to the synchronisation properties of the component dynamical systems.

While interactions are typically nonlinear in the state variables, they are often modelled as pairwise, additive and linear in the coupling weights $A_{ij}$, so that the joint effect of two nodes on a third one is the sum of the two individual (nonlinear) effects:

$$\dot{x}_i = F(x_i) + \sum_{j=1}^N A_{ij} G(x_i, x_j) \quad [2]$$

where $G$ is a coupling function describing the interactions between nodes $i$ and $j$.

### A3. Inhibition

Inhibition can take different (scale-dependent) meanings and roles (Northoff, 2002). In a *behavioural sense*, inhibition may imply suppression of ongoing behaviour or emotions. Inhibition may also be understood in terms of brain *connectivity*. A given brain region may for instance lead to inhibition of activity in another area and a lesion of the former may lead to increased activity of the latter. Finally, inhibition may be understood in a *neuronal* sense as opposed to excitation. GABA$_A$ receptors allow chloride ions to flow into the cell, thus hyperpolarizing the neuron and inhibiting neuronal firing. Notably, the former two types of inhibition do not necessarily require inhibitory neurotransmitters and may potentially be mediated by glutamatergic or other transmitter systems. How should inhibition be handled in large scale networks, particularly at mesoscales?

### A4. Higher-order structures

- Neural coupling may not be linear, sot that equation [2] (A2) may be replaced by:

$$\dot{x}_i = F(x_i) + \sum_{j=1}^{N} A_{ij} G_i(x_i, x_j) + \sum_{j,k=1}^{N} A_{ijk} G_i^{(3)}(x_i, x_j, x_k) + \cdots \quad [3]$$

where the coupling term $G_i^{(3)}$ and the corresponding coefficients $A_{ijk}$ are associated with higher-order interactions.

- A *hypergraph* $\mathcal{H} = (V, S)$ is a combinatorial object generalising ordinary graphs $\mathcal{N} = (V, E)$, where $S$ is a set contains nonempty subsets of various cardinalities of elements of $V$, called *hyperlinks*, which can connect more than two nodes (Ghoshal et al., 2009).
- An *abstract simplicial complex* is a particular hypergraph in which the set of hyperlinks is closed under inclusion, so that if a set $X$ belongs to $S$ then any subset of $X$ also belongs to $S$ (Barbarossa and Sardellitti, 2020). In an abstract simplicial complex, node-based incidence matrices are replaced by appropriate ones corresponding to boundary operators between interactions of orders differing by one.
- A *geometric simplicial complex* is the geometric counterpart of an abstract simplicial complex. A geometric simplicial complex is a pair $(X, S)$ where $X$ is a *topological space* and $S$ is a collection of continuous functions (Hatcher, 2001). It is formed by combining *simplices* in a way that satisfies two conditions: the intersection of any two simplices is a face of both, and every face of a simplex is also part of the complex. The dynamical state of the system is specified by *topological spinor* $\Psi$. For n=3, $\Psi = \begin{pmatrix} \chi \\ \psi \\ \xi \end{pmatrix}$ where entries are respectively specified on nodes, links, and triangles (Millán et al., 2025).
- A *simplex* is a particular *polytope*, i.e. a generalisation of 3D polyhedrons to any number of dimensions. An $n$-simplex is a structured set composed of points, line segments, triangles, and their n-dimensional counterparts constituting the convex hull of (i.e. the smallest convex shape containing) n+1 nodes which do not lie in any (n−1)-dimensional plane, which are glued to each other along their faces. As standard graphs, simplexes can be endowed with a direction and a set of weights.
- A *cell complex* is a structure similar to that of a simplex but which is not constrained to respect the inclusion property i.e. its subsets do not necessarily belong to the complex (Sardellitti et al., 2021).
- A *chain complex* is an algebraic structure that consists of a sequence of abelian groups and a sequence of homomorphisms between consecutive groups such that the image of each homomorphism is contained in the kernel of the next.
- *Homology groups* quantify the number of independent cycles (or "holes") of a given dimension within a topological space, which are not boundaries of higher-dimensional objects. Homology groups are topological invariants that can be used to distinguish topologically inequivalent spaces. In particular, the dimension of the first homology group $H_1$ counts the number of holes. Higher order homology groups ($H_1, H_2, \cdots$) count higher-dimensional holes. Loosely, homology quantify the extent to which a chain fails to be exact, i.e. the extent to which the image of one morphism equals the kernel of the next. A *homology class* is a finite linear combination of geometric objects with zero boundary. Each homology class is an equivalence class over cycles. Cycles in the same homology class are said to be *homologous*.
- *Simplicial homology* is a particular homology group which quantifies the number of holes of a given dimension in a simplicial complex. Simplicial homology $H_*(T)$, on a simplicial complex $T$ is constructed by triangulating a topological space $X$. $H_*(T)$ is invariant with respect to triangulation and is preserved under continuous deformations i.e. it is a *topological invariant* of $X$.
- *Betti numbers* count the number of *holes* of a given dimension on a topological surface.
- The *network Hodge Laplacian* is a generalisation of the network Laplacian and plays a crucial role in understanding the geometry and topology of manifolds. The *topological Dirac operator* is in essence a shift operator acting on spinors (Bianconi, 2021). It projects topological signals defined on one level (e.g., on nodes) to the next level (e.g., on links), allowing interactions between different dimensional elements.
- Simplicial complexes can be projected into *homological scaffolds*, which are weighted graphs based on the topology of the underlying simplicial complexes containing information about the system's hierarchical organisation (Petri et al., 2014).
- In *triadic interactions* a node regulates the interaction between two other nodes. In a hypergraph, a node may regulate the strength of a hyperlink. This is in essence the principle underlying psychophysiological interactions (Friston et al., 1997).

### A5. Generalising link types

- A *network-of-networks* is a class of structures accounting for networks interacting in various ways with other networks.
- *Multilayer networks* represent systems with multiple types of interactions, with each interaction type associated with a distinct layer and representing a distinct type of relationship between entities.
- *Multiplex networks* are a particular kind of multilayer network in which inter-layer links are restricted to the projections of the same node at different layers.
- In a *multilayer network*, triadic interactions mediate inter-layer node interactions. Neural networks and networks of glia cells may form two layers of a multiplex network interacting via triadic interactions.
- *Annotated networks* are networks equipped with additional data or metadata describing the properties of the nodes or links. Annotations allow encoding different types of entities and relationships (and interactions) (Newman and Clauset, 2016).

### A6. Beyond single networks

- A *network ensemble* is a probability distribution on graphs. Specifically, a network ensemble is a family of networks that satisfy a set of constraints, e.g. a given number of nodes and links, or degree distribution. The role of a given structural characteristic in shaping the network can be quantified by the ensemble's entropy, i.e. the normalised logarithm of the number of networks in the ensemble (Bianconi, 2007, 2009). *Random graph theory* studies the asymptotic behaviour of such ensembles as the number of nodes $N \to \infty$ and the connection probability $p = p(N)$, and a property is said to happen with high probability if the probability approaches one as $N \to \infty$ (Kahle, 2014).
- The *path integral approach* describes the evolution of a system, e.g. its correlation function, as the weighted sum over all possible paths it can take between two points in space and time, each of which is assigned a probability. This can for instance be done by deriving a generating functional for the relevant correlation and response functions induced by the dynamics (Crisanti and Sompolinsky, 2018).
- *Bag-of-paths* considers all possible paths in a network as a set of independent elements, defining a probability distribution over these paths, which defines the relatedness and generalised distances within the network (Françoisse et al., 2017).
- In a cellular automaton, a discrete model of computation evolving in time according to certain rules, *path diversity* $\mathcal{D}$ counts the number of nonequivalent paths from an attractor to a transient state with a configuration that cannot be produced from a previous configuration, following the automaton's rules (Shreim et al., 2007).
- *Network density matrices* describe the statistical state of a netwok by considering that each node or component is in a quantum state, i.e. a superposition of multiple states (Ghavasieh and De Domenico, 2022).
- *Graphons* are the limiting object for sequences of finite graphs, which capture essential features of a graph's structure as the number of vertices grows (Lovász and Szegedy, 2006).

## A7. Effective field theories and the renormalisation group approach

- An *effective field theory* describes physical phenomena occurring at a given length (or energy scale), ignoring substructure and degrees of freedom at shorter lengths (or higher energies) (Georgi, 1993). This involves averaging over the behaviour of the underlying theory at shorter length scales to derive a simplified model at longer length scales including the appropriate degrees of freedom. Effective field theories typically work best in the presence of a large separation between length or time scale of interest and that of the underlying dynamics, an assumption not always fulfilled by brain dynamics (Papo, 2014a).
- A model is said to be *renormalisable* if the changes caused by modification of microscopic parameters can be summarised by a finite number of phenomenological parameters. *Renormalisation* reflects the idea that small-scale details average out at large enough spatial and temporal scales.
- The *renormalisation group* approach allows describing systems with many degrees of freedom across different levels of resolution (Kadanoff, 1971; Wilson and Kogut, 1974). The *renormalisation group flow* defines transformations and coarse-graining schemes to average overs small scale details and ultimately define effective degrees of freedom and their interactions at a given scale. Renormalisation can operate in real, conjugate, phase space or in time. The renormalisation flow is in essence a generalised dynamical system where the rescaling factor (or number of iterations) plays the role of time. Asymptotic behaviour of renormalisation may in principle converge towards any kind of attractor, in the simplest case, a hyperbolic fixed point. Fixed points are associated with the system's critical exponents. Thes stable and unstable manifolds partition the space of models into universality classes, the stable manifold representing basins of attraction of probability distribution functions (Jona-Lasinio, 2001). Thus, the renormalisation group approach allows characterising systems with similar large-scale properties into *universality classes*. It can also be used to detect systems' phase transition points and its behaviour around them. A system may exhibit many possible asymptotic behaviours and the particular one attained by the system under coarse graining depends on the physical parameters' initial values and their location in the fixed points' basins of attraction. For instance, power-law behaviour at a given scale may evolve to a domain where the system is characterised by hierarchical behaviour (Pérez-Mercader, 2004).
- The renormalisation process comprises three main steps: *coarse-graining*, averaging out of fine details, and coupling and parameter rescaling (Lesne, 2008a). A fourth step is needed when considering bare brain anatomy or dynamics as continuous fields, as this continuous-to-discrete mapping necessarily implies infinities.
- *Renormalisation operators* are functions that perform these scale transformations. A renormalisation operator is said to be *relevant* if its coupling constants grow with the flow, indicating that the operator becomes more important at larger scales (or lower energies), and *irrelevant* if its coefficients decrease as the energy scale is lowered, meaning they become less important at lower energies. Relevant directions control the scaling exponents' value, while irrelevant ones only provide corrections to scaling. Most systems' macroscopic physics is dominated by only a few observables, as most observables are irrelevant. The differences among the fine-scale components across systems are determined by irrelevant observables, while the relevant observables are shared by many systems which may be profoundly different at shorter lengths. This allows grouping macroscopic phenomena into a small set of universality classes, specified by the shared sets of relevant observables.
- In its standard form, the renormalisation group approach is predicated upon the notions of homogeneity, symmetry and locality. Biological networks such as the brain generally lack all of these properties but can nonetheless be renormalised in both real and conjugate space (Song et al., 2006; Gfeller and de los Rios, 2007; Radicchi et al., 2009; Rozenfeld et al., 2010; Aygün and Erzan, 2011; Bradde and Bialek, 2017; García-Pérez et al., 2018; Garuccio et al., 2023; Villegas et al., 2023; Gabrielli et al., 2025).

## A8. Category theory and functoriality

- A *category* $C$ consists of a class of objects of the same type, and a class of maps between these objects, called *morphisms*, which contains the identity mappings and is closed with respect to mapping composition.
- A *functor* is a morphism between two categories which preserves the structure and relationships between objects and morphisms. Thus, a functor respects both the identity morphisms and the composition of morphisms in the original category when mapping them to the target category.
- A mapping is *functorial* if it preserves composition and identities. *Functoriality* expresses the idea that a *functor* $F: C \to D$ must preserve the inherent structure of the original category $C$ when mapping it to category $D$. For instance, considering a continuous map $f: X \to Y$ between simplicial complexes, in the same way that $X$ induces a chain complex, $f$ induces a chain map, i.e. a sequence of homomorphims from $C_m(X) \to C_k(Y)$. Together with the boundary maps this chain map forms a *commutative diagram*, i.e. a graphical representation depicting how composite morphisms relate to each other diagram wherein all directed paths with the same start and endpoints lead to the same result (Ghrist, 2014). Specifically, a diagram is *commutative* if all possible paths in the diagram corresponding to compositions of morphisms, represent the same function or relationship.
- *Self-dissimilarity* quantifies the extent to which a system's structure observed at different scales differ from each other as the amount of extra information required to describe a system on one scale, given a description on another scale (Wolpert and MacReady, 2000; Itzkovitz et al., 2005).

## A9. Emergence, network structure, and function

- *Emergence* designates properties or behaviour that are different from those of its individual components, and which arise from the interactions among the system's parts. Emergent properties of a system may not be stable or consistent across all scales or transitions (Varley, 2022). Even though a system may exhibit emergent behaviour, its local dynamics can be unpredictable or even non-emergent.
- The *slaving principle* states that near instabilities, complex systems' macroscopic behaviour is dominated by a few slow-varying variables (collectively termed *order parameter*), which control the behaviour of much faster variables.
- The *maximum entropy principle* (Jaynes, 1957) aims to find the probability distribution that best represents the data while making the least assumptions beyond what is explicitly given in the data. The principle states that the most appropriate distribution to model a given dataset is the one with the highest entropy, subject to the constraints imposed by the data.
- *Generative models* are algorithms that learn the underlying probability distribution of a dataset and can be used to generate new samples from that distribution. Generative models can be used to simulate or generate plausible neuronal dynamics (at multiple scales) or to make inferences about the functional form and architecture of distributed neuronal processing (Vértes et al., 2012; Betzel and Bassett, 2017; Medrano et al., 2024). The generative model is used as an observation model and optimised to best explain some data. Crucially, this optimisation entails identifying both the parameters of the generative model and its structure, respectively via model inversion and selection.

- Macro-scale dynamics can have stronger causal effects than micro-scale dynamics, a notion termed *causal emergence* (Hoel et al., 2013). *Effective information*, a measure quantifying the strength of causal interactions between parts of a system identifies the scales where causal relationships are most pronounced (Hoel et al., 2013).
- *Network geometry with flavour* allows characterising network geometry in any dimension, by using some non-equilibrium dynamical process to evolve simplicial complexes (Bianconi and Rahmede, 2015, 2016). The process can generate various discrete geometries, e.g. higher-dimensional manifolds, and scale-free networks. This structure can be equipped with a *flavour*, i.e. a parameter that can change the topological nature of the simplicial complex and its evolution. Different values of the flavour parameter can lead to the emergence of different network topologies.
- The *nerve of a cover* is a simplicial complex constructed from an open cover (i.e. a collection of open subsets of a given set whose union contains the set) by taking the intersections of the open sets composing the cover. The nerve captures the topological properties of the original space using a discrete, combinatorial representation. The *nerve theorem* ensures that if the cover is sufficiently fine, the nerve is *homotopy equivalent* to the original space, i.e. it can be continuously deformed into that space, preserving its topological characteristics. *Čech cohomology* provides a way to characterise global topological properties based on the intersection properties of its open covers. The intersections of the open sets induce a structure e.g. a simplicial complex, that captures how the open sets are related. This complex is used to compute *cohomology groups*, which encode topological properties such the number of connected components, or twists in the surface (Ghrist, 2014).

**A10. Brain function, non-trivial properties and exotic spaces**

- *Degeneracy* refers to a situation where multiple solutions or parameters can achieve the same outcome.
- *Sloppiness* describes systems where many parameter values produce nearly identical results.
- *Orbifolds* are differentiable manifolds containing singularities. Orbifolds are topological spaces which locally resemble the quotient space of a Euclidean space under the linear action of a finite group.
- *Pretopological spaces* generalise topological spaces by relaxing the restrictions on closure operators. This allows studying structures where "closeness" might not be as tightly defined as in standard topology.

**Funding statement**

JMB acknowledges support from Ministerio de Ciencia e Innovación under grants PID2020-113737GB-I00 and PID2023-147827NB-I00.

**Conflict of interest statement**

The authors have no conflict of interest to declare.